\PassOptionsToPackage{usenames,dvipsnames}{xcolor}
\documentclass[conference]{IEEEtran}
\IEEEoverridecommandlockouts

\usepackage{fancyhdr}

\fancypagestyle{IEEEtitlepagestyle}{
    \fancyhf{}
    
    \fancyhead[C]{\footnotesize
    2026 IEEE 33rd International Conference on Electronics, Circuits and Systems (ICECS)}
    
    \fancyfoot[L]{\footnotesize
    979-8-3195-1905-4/26/\$31.00~\copyright~2026 IEEE}

}

\usepackage{cite}
\usepackage{amsmath,amssymb,amsfonts}
\usepackage{algorithmic}
\usepackage{graphicx}
\usepackage{textcomp}
\usepackage[usenames,dvipsnames]{xcolor}
\usepackage{subcaption}
\usepackage{url}
\usepackage{float}
\usepackage[normalem]{ulem}
\usepackage{booktabs}
\usepackage[strings]{underscore}
\usepackage{enumitem}
\usepackage[numbers]{natbib}
\usepackage[switch,columnwise]{lineno}
\usepackage{mathabx}
\usepackage{pifont}
\usepackage{dirtytalk}
\usepackage[switch,columnwise]{lineno}

\def\BibTeX{{\rm B\kern-.05em{\sc i\kern-.025em b}\kern-.08em
    T\kern-.1667em\lower.7ex\hbox{E}\kern-.125emX}}

\newcommand{\TROJAN}{\mbox{\textsc{IRT}}}

\begin{document}

\bstctlcite{IEEEexample:BSTcontrol}

\title{
Assessing Runtime Electromagnetic Detection of CPU Hardware Trojans Targeting Kernel Memory
}

\makeatletter
\newcommand{\linebreakand}{%
  \end{@IEEEauthorhalign}
  \hfill\mbox{}\par
  \mbox{}\hfill\begin{@IEEEauthorhalign}
}
\makeatother


\author{
\IEEEauthorblockN{
Athanasios Moschos\IEEEauthorrefmark{1},
Baki Berkay Yilmaz\IEEEauthorrefmark{2},
Kevin Valakuzhy\IEEEauthorrefmark{1},
Angelos D. Keromytis\IEEEauthorrefmark{1}
}

\IEEEauthorblockA{\IEEEauthorrefmark{1}
Georgia Institute of Technology,
School of Electrical and Computer Engineering, USA}

\IEEEauthorblockA{\IEEEauthorrefmark{2}
Wayne State University,
Department of Electrical and Computer Engineering, USA}
}

\IEEEaftertitletext{%
    \centering
    \footnotesize
    \parbox{0.96\textwidth}{%
        \centering
        \textcopyright~2026 IEEE.
        Personal use of this material is permitted.
        Permission from IEEE must be obtained for all other uses,
        in any current or future media, including
        reprinting/republishing this material for advertising or
        promotional purposes, creating new collective works, for
        resale or redistribution to servers or lists, or reuse of
        any copyrighted component of this work in other works.
    }%
    \vspace{0.5\baselineskip}
}

\maketitle

\begin{abstract}
    Side-channels are a promising approach for hardware trojan detection, as they can passively reveal malicious hardware activity without requiring additional circuitry or destructive analysis of the device under test.
This work investigates the use of electromagnetic (EM) emanations for runtime detection of hardware trojan attacks.
Using an open-source hardware trojan that implements an arbitrary memory-write primitive, we tamper with the kernel memory of a Linux system running on a RISC-V microarchitecture and perform a real-time baseline detection using the processor's EM emissions.
Our results indicate that, under certain conditions, hardware trojans can be detected indirectly through the anomalous software behavior they enable.
\end{abstract}

\begin{IEEEkeywords}
Hardware Trojans, Signal Processing, Operating Systems, Computer Architecture, Electromagnetic Side-channels
\end{IEEEkeywords}

\section{Introduction}
\label{sec:introduction}
    \par Detecting hardware trojans during chip deployment is becoming increasingly important due to the absence of industry-wide standards for identifying malicious hardware logic prior to fabrication.
This challenge is especially pronounced in CPUs, whose complex operation and diverse microarchitectural activity produce highly variable execution patterns.
In this work, we investigate the runtime detection of hardware trojans that target the dynamically allocated kernel memory within CPUs.
Kernel memory is particularly important as it hosts security-relevant runtime objects (\textit{e.g.,} task credentials, process descriptors, security policy structures).
We show that electromagnetic emissions encode information about the accessed memory region, kernel or user, allowing us to distinguish malicious activity on dynamically allocated kernel memory.
Building on this observation, we develop a golden-free runtime detection method that targets such trojan-assisted activity.
To the best of our knowledge, this approach to EM-based trojan detection has not been previously explored.

\par Prior work~\cite{CRZAPM2014, nazari2017eddie} demonstrated that software-level asymmetries can produce distinct electromagnetic signatures.
Specifically,~\citet{CRZAPM2014} showed how electromagnetic measurements can identify instruction-level execution events.
Deviation from expected code-execution patterns is also identifiable~\cite{nazari2017eddie}.
Other EM methods, such as backscattering~\cite{NLNYBBPMZA2020}, identify hardware logic modifications directly, but require physical access to the chip surface.
In practice, this requirement is often difficult to satisfy for deployed CPUs due to standard packaging and cooling components, such as heatsinks.




\section{Tampering With Kernel Memory}
\label{sec:hw_tj_attack}
    \paragraph*{Hardware Trojan Design}
We use one of the open-source Interrupt-Resilient Trojan designs~\cite{MAMFKAD2024}, \TROJAN-1, that features a trigger mechanism implemented within the processor's register file.
The trojan's payload circuit is designed to bypass the operating-system-enforced isolation between privileged and non-privileged memory regions.
Typically, kernel pages have their user-mode bit cleared to prevent user-level accesses.
To enable malicious overwrites from user space, the payload circuit masks the user-mode bit of kernel page table entries (PTEs), thereby suppressing memory-access exceptions.

\paragraph*{Attack Emulation}
Using a custom Linux kernel module (LKM), we allocate a 4~MB integer array within the kernel space and expose its base address to the user space, similar to a memory-disclosure primitive.
Array allocation employs either \texttt{kmalloc} or \texttt{vmalloc}.
A malicious user-space binary can tamper with kernel memory by targeting the disclosed address.

\par To assess the spectral difference between malicious kernel-memory overwrites and benign user-memory overwrites, the binary allocates a separate 4~MB user-space array through \texttt{malloc} and generates random in-bounds indices.
At runtime, the binary's overwrite routine targets either the user- or the kernel-space array using those indices.
Hence, both scenarios have the \textit{exact same execution flow}.
Activating the payload only during the overwrite routine bounds the interval of suppressed memory-access exceptions.

\section{EM-Based Detection Methodology}
\label{sec:method}
    \par To assess the sensitivity of EM-based monitoring to trojan activity, we adapt the supervised-learning methodology of~\cite{yilmaz2019detecting} to a one-class classification approach.
Our method comprises three stages.
Initially, during a two-stage \textit{training process}, EM activity from various CPU operations is characterized to construct a model of benign EM signatures.
In the subsequent \textit{testing stage}, EM signals captured from live CPU operation are processed and compared against this training model in real time to identify anomalous activity.

\paragraph*{Training Stage-1}
We acquire long EM waveforms, which we partition into shorter time segments, each with a duration equal to the \textit{sample time}. 
Using a preconfigured \textit{window size} and \textit{overlap ratio}, we compute the Short-Time Fourier Transform (STFT) for each segment.
We combine the resulting sequence of time-frequency representations to construct a two-dimensional array analogous to a spectrogram, with one axis being the frequency bins and the other the time.

\par Next, we apply max-pooling to this time-frequency array, motivated by two considerations.
First, we expect trojan activity to primarily affect dominant frequencies of the acquired signal.
The monitoring system is constantly capturing emanations, unaware of the exact trojan activation interval.
Hence, max-pooling preserves the effects of short-lived anomalies appearing in these dominant frequencies.
Second, the spectral content at certain frequencies can be unstable, exhibiting smearing across neighboring frequencies.
Such behavior could mislead the distance-based classifier of the testing stage, due to the accumulation of small differences across the non-dominant frequency bands.
Consequently, max-pooling provides a more compact and robust spectral representation, through a lower-dimensional time-frequency array that preserves anomalous activity in dominant frequencies and reduces sensitivity to local smearing-induced spectral fluctuations. 

\par Another consideration is noise present in the measurement setup.
To mitigate noise, we average the rows of the reduced-dimensionality array, producing a single spectral representation.
We convert this representation to dB to allow the distance-based classifier to also capture the relative changes occurring in weaker spectral regions.
Fluctuations in the global power level across measurements of the same CPU operation can increase variability.
We account for this by applying a normalization scheme: we subtract from each single spectral representation its mean and subsequently divide it by its maximum magnitude.
Hence, we preserve the overall spectral shape and minimize the effect of global power fluctuations.

\par Overall, the training stage-1 pipeline is computationally lightweight, and is used both to construct the training-stage EM-signature model and to process the real-time emanations at the testing stage.
Consequently, computations in this stage must complete within the waveform acquisition intervals.

\paragraph*{Training Stage-2}
This stage contributes only to the EM-signature model generation.
The stage-1 measurements correspond to distinct CPU operations (\textit{e.g.,} memory accessing), each one with multiple spectral representations.
Since normalization rescaled every spectral representation to a comparable range, we apply the $k$-means algorithm to reduce the size of each EM model associated with a CPU operation. 
The resulting cluster centroids of each EM model compactly represent the dominant CPU activity patterns, forming a unified EM-signature database for efficient inference during testing.

\paragraph*{Testing Stage}
To classify whether live EM signals are the outcome of anomalous activity, we compute the $\ell_1$-norm distance between each acquired signal (after stage-1 processing) and the database's EM-signatures.
The smallest resulting distance is then compared against a predefined threshold, flagging the signal as anomalous if the threshold is exceeded.

\section{Experimental Setup}
\label{sec:experimental_setup}
    \begin{table}[!t]
\caption{Resource utilization of CVA6 and \TROJAN-1.}
\scriptsize
\centering
\scalebox{1}{
\begin{tabular}{*{5}{c}}
    \toprule
    \textbf{\shortstack{Module\\Name}}& \textbf{\shortstack{Design\\LUTs}}& \textbf{\shortstack{Design\\FFs}}& \textbf{\shortstack{Design-to-CVA6\\LUT Ratio (\%)}}& \textbf{\shortstack{Design-to-CVA6\\FF Ratio (\%)}}\\
    \midrule
    CVA6 w. \TROJAN-1& 72,371& 46,878& 100& 100\\
    \TROJAN-1 Trigger& 26& 1& 0.0359& 0.0021\\
    \TROJAN-1 Payload& 4& 3& 0.0055& 0.0064\\
    \bottomrule
\end{tabular}
}
\label{table:fpga_utilization}
\vspace{-2em}
\end{table}

\paragraph*{Trojan Testbed}
Similar to~\cite{MAMFKAD2024}, we use the 64-bit CVA6 microarchitecture~\cite{FZLB2019} to implement the \TROJAN-1~trojan.
The experiments are performed on the Genesys 2 board, with the trojan-afflicted CVA6 operating at a 50~MHz clock and booting a Linux system based on kernel v5.10.7.

\par The ratio of the malicious logic to the overall size of the trojan-afflicted CVA6 implementation is presented in Table~\ref{table:fpga_utilization}.
Higher ratios yield increased chances of discovering the trojan through its side-channel emanations \cite{NLNCCLPMZA2019}. 
In our case, the very small trojan-to-host-design ratio in the generated bitstream challenges our detection approach~\cite{BSHMSBMNS2014}.

\paragraph*{Measurement Setup}
Our monitoring system uses a TekBox 20~mm near-field magnetic probe to measure EM emanations during CVA6's operation.
The probe is connected to a B205mini-i SDR for signal acquisition, with an Apple M4 MacBook serving as the real-time processing platform.
Similar to~\cite{NLNYBBPMZA2020}, we measure frequency ranges around the CVA6 clock harmonics.
To determine the optimal probe location and center frequency, we manually search for a location beneath the FPGA with strong EM emissions during the execution of memory operations.
We find a favorable center frequency at 265~MHz, near the 5th clock harmonic.
The SDR's bandwidth is set to 8~MHz to minimize data throughput and the processing overhead of runtime measurements.
Our method is golden-free, hence the trojan-afflicted CVA6 is used to capture measurements during both the training and the testing stage.

\section{Results}
\label{sec:results}
    \subsection{Spectral Representation of Memory Overwrites}
\par To illustrate the spectral characteristics associated with overwrites in the kernel- and user-space memory, we collect EM signals during five distinct scenarios spanning legitimate and malicious activity.
We pass the measurements through the training stage-1 processing before plotting their spectrograms.

\paragraph*{Kernel Overwrites}
We modify the LKM described in Section~\ref{sec:hw_tj_attack} to allocate a 4~MB integer array either via \texttt{kmalloc} or \texttt{vmalloc}, and subsequently perform 300~K back-to-back element overwrites at random indices.
The spectral characteristics of this procedure are shown in Fig.~\ref{fig:kernel_activity}.
Notably, overwrites on the \texttt{vmalloc}-allocated object exhibit a significantly stronger EM signature, while taking longer to execute.
This temporal and spectral divergence is attributed to differences in the page-table-walking activity.

\par These two memory-allocation functions rely on inherently distinct virtual-to-physical memory-mapping mechanisms.
Regions allocated through \texttt{vmalloc} face a substantially higher address-translation overhead, as \texttt{vmalloc} constructs a contiguous virtual region of 4~KB physical pages. 
For a 4~MB-sized array, this results in multiple leaf page table entries.
In contrast, \texttt{kmalloc} allocates physically contiguous memory from regions covered by the kernel's direct mapping of RAM.
Since this mapping is typically backed by huge pages~\cite{JC2022}, an array of equivalent size is covered by only two 2~MB pages, resulting in more efficient memory accesses.

\begin{figure}[!t]
    \caption{Spectrograms of random kernel overwrites in a 4~MB array allocated via \texttt{kmalloc} (\textit{top}) and \texttt{vmalloc} (\textit{bottom}).}
    \label{fig:kernel_activity}
    \centering
    \includegraphics[width=\linewidth]{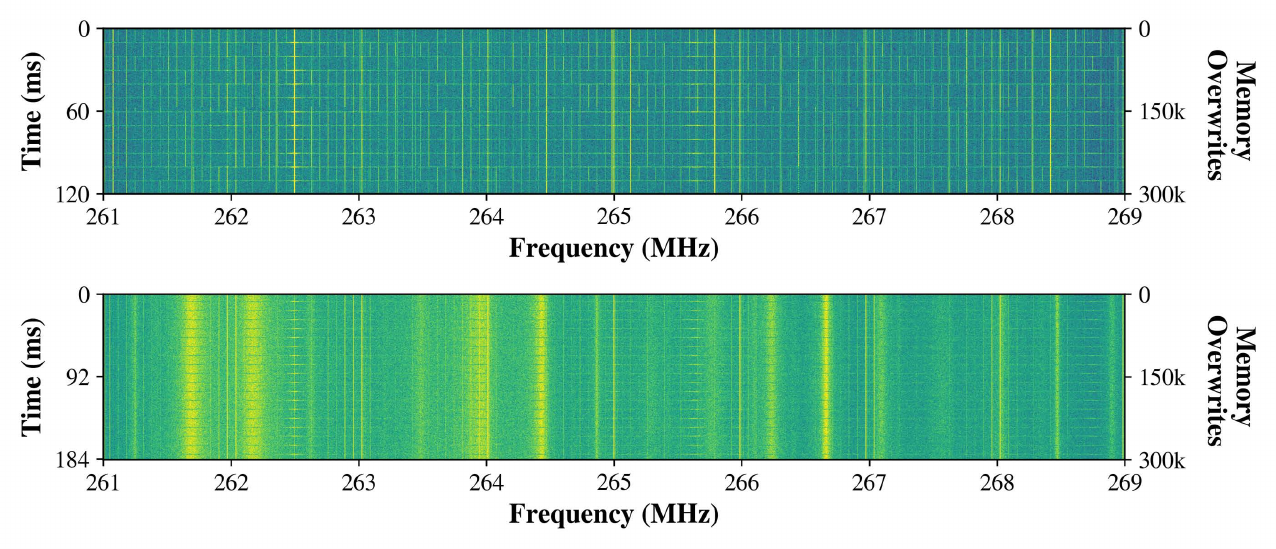}
    \vspace{-1em}
\end{figure}

\begin{figure}[!t]
    \caption{Spectrograms demonstrating a user process performing benign and trojan-assisted memory overwrites.
    The \textit{top} shows benign random overwrites on a 4~MB user-space region allocated with \texttt{malloc}. 
    The \textit{middle} and \textit{bottom} spectrograms show malicious random overwrites on a 4~MB kernel-space region allocated with \texttt{vmalloc} and \texttt{kmalloc}, respectively.
    }
    \label{fig:user_activity}
    \centering
    \includegraphics[width=\linewidth]{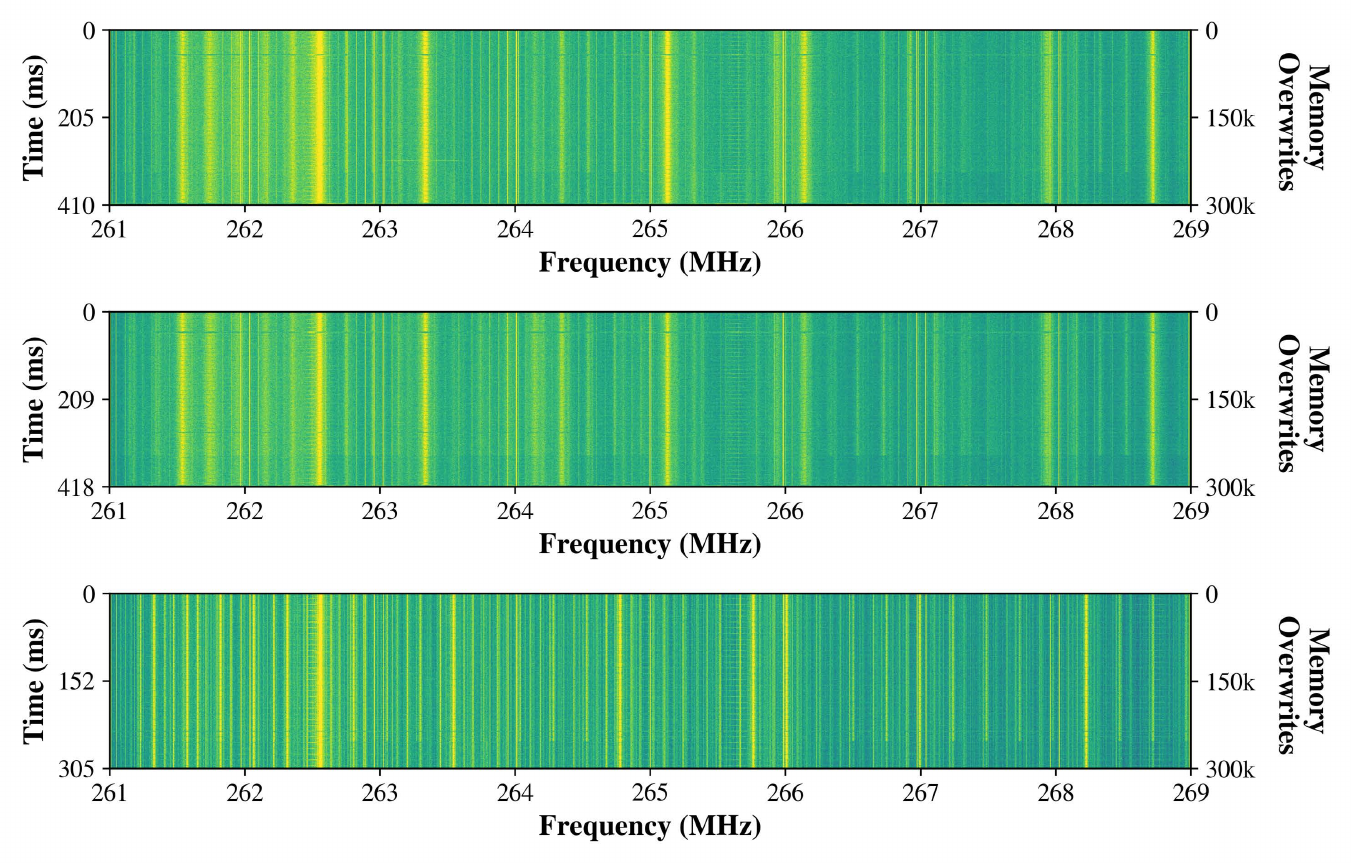}
    \vspace{-2em}
\end{figure}


\paragraph*{User Process Overwrites}
For this experiment we use the LKM and the user-space binary described in Section~\ref{sec:hw_tj_attack}.
We capture measurements while the binary performs benign and malicious overwrites in the allocated arrays.
During the overwrite routine, \TROJAN-1 is activated and the exact same code path executes in all cases.
The spectrograms for these scenarios can be seen in Fig.~\ref{fig:user_activity}.
The spectral and temporal differences between the top (\texttt{malloc}) and middle (\texttt{vmalloc}) spectrograms are insignificant, as large user-space allocations obtained through \texttt{malloc} are also backed by conventional 4~KB pages.
This results in similar demand for address translations and significant page-table-walking activity.

\par In contrast, the bottom spectrogram (\texttt{kmalloc}) presents temporal and spectral discrepancies distinguishable from the other two.
This variation is not the outcome of the malicious hardware switching activity, but rather the absence of the page-table-walking activity during the malicious overwrites.

\par We must note here that pages larger than 4~KB can also be allocated to user-space processes via the kernel feature of transparent huge pages or THP.
This could result in fewer address translations, as in the case of \texttt{kmalloc}.
However, THP support for RISC-V was introduced after Linux v5.14, and is unavailable in our kernel version.
Moreover, THP behavior is dependent on the kernel configuration and operating system distribution.
Thus, it is outside the scope of our evaluation.




\subsection{Spectrum-Based Monitoring Evaluation}
\begin{figure*}[!t]
    \caption{Heatmap evaluating detection capability of distinct training models under varying trojan-assisted memory overwrites.}
    \label{fig:heatmap}
    \centering
    \includegraphics[width=0.9\textwidth]{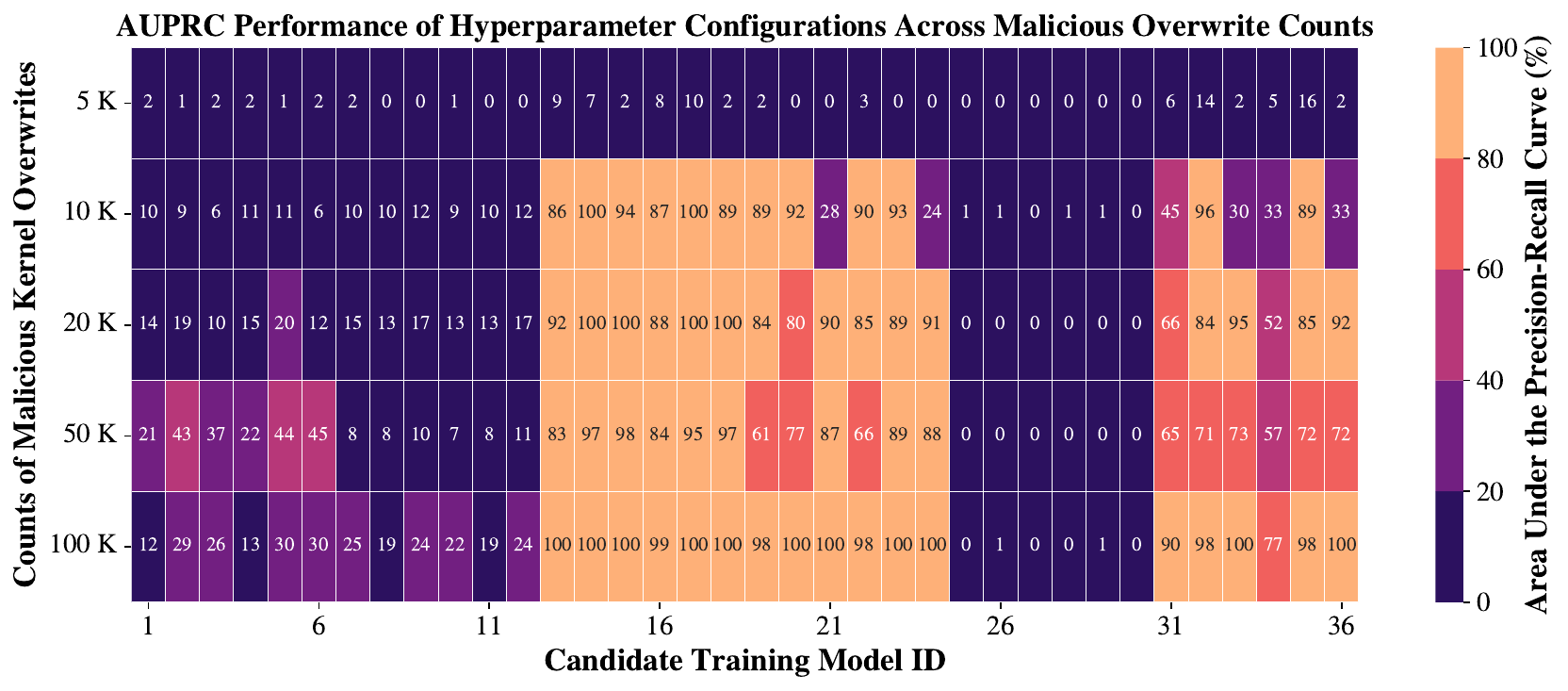}
\end{figure*}

\begin{table}[!t]
\caption{Hyperparameter configurations per training model.}
\scriptsize
\centering
\scalebox{1}{
\begin{tabular}{*{5}{c}}
    \toprule
    \textbf{\shortstack{Training\\Model ID}}& \textbf{\shortstack{Size of STFT\\Window}}& \textbf{\shortstack{Overlap\\Ratio}}& \textbf{\shortstack{Max-Pooling\\Kernel Size}}& \textbf{\shortstack{Sample\\Time (ms)}}\\
    \toprule
    $[1, 2, 3]$ & 1024 & 0.1 & 4 & $[5, 10, 20]$\\
    $[4, 5, 6]$ & 1024 & 0.5 & 4 & $[5, 10, 20]$\\
    $[7, 8, 9]$ & 4096 & 0.1 & 4 & $[5, 10, 20]$\\
    $[10, 11, 12]$ & 4096 & 0.5 & 4 & $[5, 10, 20]$\\
    $[13, 14, 15]$ & 1024 & 0.1 & 32 & $[5, 10, 20]$\\
    $[16, 17, 18]$ & 1024 & 0.5 & 32 & $[5, 10, 20]$\\
    $[19, 20, 21]$ & 4096 & 0.1 & 32 & $[5, 10, 20]$\\
    $[22, 23, 24]$ & 4096 & 0.5 & 32 & $[5, 10, 20]$\\
    $[25, 26, 27]$ & 1024 & 0.1 & 64 & $[5, 10, 20]$\\
    $[28, 29, 30]$ & 1024 & 0.5 & 64 & $[5, 10, 20]$\\
    $[31, 32, 33]$ & 4096 & 0.1 & 64 & $[5, 10, 20]$\\
    $[34, 35, 36]$ & 4096 & 0.5 & 64 & $[5, 10, 20]$\\
    \bottomrule
\end{tabular}
}
\label{table:training_models}
\vspace{-2em}
\end{table}

Our training model incorporates the spectral representations shown in Fig.~\ref{fig:kernel_activity} and the user-space overwrite activity of Fig.~\ref{fig:user_activity} (top).  
The malicious activity (middle \& bottom) in Fig.~\ref{fig:user_activity} is the target behavior that our monitoring system aims to detect during testing and flag it as kernel tampering.
The $\ell_1$-norm distance metric can reliably distinguish successive overwrites of \texttt{kmalloc}-backed kernel objects, owing to their distinct spectral signature.
However, the spectral activity generated by successive overwrites of \texttt{vmalloc}-allocated objects closely resembles the emissions of benign user-space overwrites, making such tampering events hard to detect with our system.

\par In practice, the attack duration is not known a priori.
A live monitoring system must therefore detect EM anomalies across attacks of varying duration.
We approximate this variability by progressively reducing the number of malicious kernel-space overwrites on \texttt{kmalloc}-allocated memory and evaluating the resulting detection sensitivity across candidate training models.
For this quantitative assessment, we use the area under the precision-recall curve (AUPRC) and summarize the results in the heatmap of Fig.~\ref{fig:heatmap}.
We choose AUPRC due to the dataset imbalance between trojan and benign activity.
Table~\ref{table:training_models} presents the hyperparameters used to construct the candidate models.

\par Models 13--24 consistently perform best across malicious overwrite counts from 100~K down to 10~K, with models 14 and 17 achieving $100\%$ AUPRC even at 10~K; only models 21 and 24 show notably degraded performance at this count.
These results indicate that the max-pooling kernel size strongly influences detection performance, with the medium-sized value of 32 yielding the best overall results.
Models 32 and 35 also achieve comparable AUPRC at 10~K, while using a larger STFT window size of 4096 samples and a large kernel size of 64.
Notably, a 10~ms sample time is common to the high-performing models 14, 17, 32 and 35.

\par Nevertheless, shorter overwrite sequences produce less distinctive spectral patterns, as reflected by the very low AUPRC scores across all models at 5~K overwrites.
An attacker who distributes memory modifications across multiple short bursts may therefore evade detection.
Future work will focus on improving detection at such low overwrite counts.

\section{Discussion}
\label{sec:discussion}
    \par The very small footprint of \TROJAN-1 relative to the overall CVA6, combined with its intertwined placement within the CPU layout, complicates acquisition of a pristine trojan signature at runtime~\cite{NLNCCLPMZA2019, BSHMSBMNS2014}.
Spectrally diverse electromagnetic emissions from various microarchitectural components further exacerbate this challenge.
Nevertheless, our results indicate that trojan-assisted software execution can produce distinct EM signatures detectable without comparison against trojan-free measurements.
We therefore argue that CPU hardware trojans can be detected \textit{through the anomalous software behavior they enable}. 
This marks a departure from traditional detection approaches that seek to isolate the minute power or EM variations directly attributable to the trojan logic.

\par To understand whether this shift in detection strategy can practically aid identification of kernel-memory tampering, we examine the prevalence of different kernel memory-allocation mechanisms.
We analyzed the most recent Linux kernel source code available at the time of writing (v7.1-rc6) for common allocation APIs, including \texttt{kmalloc/vmalloc} and their variants.
Our analysis identified approximately 36$\times$ more call sites for \texttt{kmalloc}-style allocators than for \texttt{vmalloc}, with over 36,000 and roughly 1,000 call sites, respectively.
These results suggest that untargeted attacks are likely to interact with kernel objects allocated via \texttt{kmalloc}-based mechanisms and therefore exhibit detectable emissions.
Conversely, carefully scoped modifications of \texttt{vmalloc}-backed objects may evade detection, compromising system security.
Consequently, favoring \texttt{kmalloc}-based allocation for security-sensitive kernel objects could improve the runtime detectability of kernel-memory tampering attacks via EM side-channels.

\bibliographystyle{IEEEtranN}
\bibliography{references}

\end{document}